\documentclass[twocolumn,aps,showpacs]{revtex4}
\usepackage{color}
\usepackage{graphics}
\usepackage{epsfig}
\usepackage{amsmath}
\usepackage{amssymb}
\usepackage{amsthm}
\usepackage{amsmath,alltt}
\usepackage{graphicx}
\usepackage{amssymb}
\usepackage[latin1]{inputenc}
\usepackage[T1]{fontenc}
\usepackage{color}

\newcommand{\bea}{\begin{eqnarray}}
\newcommand{\eea}{\end{eqnarray}}
\newcommand{\beq}{\begin{eqnarray}}
\newcommand{\eeq}{\end{eqnarray}}

\newcommand{\be}{\begin{equation}}
\newcommand{\ee}{\end{equation}}

\newcommand{\VA}{\langle A^2 \rangle}
\newcommand{\eq}[1]{Eq.~(\ref{#1})}

\begin{document}
\title{QCD: Restoration of Chiral Symmetry and Deconfinement for Large $N_f$}
\author{A. Bashir$^1$, A. Raya$^1$, J. Rodr\'iguez-Quintero$^2$}
\affiliation{ $^1$Instituto de F\'isica y Matem\'aticas,
Universidad Michoacana de San Nicol\'as
de Hidalgo, Edificio C-3, Ciudad Universitaria, Morelia, Michoac\'an 58040, M\'exico.\\
$^2$ Departamento de F\'isica Aplicada, Facultad de Ciencias
Experimentales, Universidad de Huelva, Huelva 21071, Spain.}

\begin{abstract}

Exploiting the recent lattice results for the infrared gluon
propagator with light dynamical quarks, we solve the gap equation
for the quark propagator. We thus model the chiral symmetry
breaking mechanism with increasing number of flavours and study
confinement (intimately tied with the analytic properties of QCD
Schwinger functions) order parameters. We obtain, with this
approach, clear signals of chiral symmetry restoration and
deconfinement when the number of light quark flavors exceeds a
critical value of $N_f^c \approx 8 \pm 1$, in agreement with the
state-of-the-art direct lattice analysis of chiral symmetry
restoration in QCD.

\end{abstract}


\maketitle

\date{\today}


\section{Introduction}

Quantum chromodynamics (QCD) with large number of massless fermion
flavors has seen a resurgence of interest due to its connection
with technicolor models, originally proposed by Weinberg and
Susskind~\cite{Weinberg:1979}, which fall into the category of
\emph{Beyond the Standard Model Theories}. They possess
intrinsically attractive features. They do not resort to
fundamental scalars to reconcile local gauge symmetry with massive
mediators of interaction and have close resemblance with
well-studied fundamental strong interactions, i.e., QCD. However,
their simple versions do not live up to the experimental
electroweak precision constraints, in particular the ones related
to flavor changing neutral currents. Walking models containing a
conformal window and an infrared fixed point can possibly cure
this defect and become phenomenologically
viable~\cite{Holdom:1985}. This scenario motivates the
investigation of QCD for similar characteristics. One looks for
such behavior of QCD for large number of light flavors
\emph{albeit} less than the critical value where asymptotic
freedom sets in, i.e., $N_f^{c_1} = 16.5$, a Nobel prize winning
result known since the advent of QCD,~\cite{Wilczek:1973}. Just as
$N_f$ dictates the peculiar behavior of QCD in the ultraviolet, we
expect it to determine the onslaught of its emerging phenomena in
the infrared, i.e., chiral symmetry breaking and confinement.

Whereas the self interaction of gluons provides anti-screening,
the production of virtual quark-antiquark pairs screens and
debilitates the strength of this interaction of non Abelian
origin. For real QCD, light flavors are small in number and hence
yield to the gluonic influence which triggers confinement and
chiral symmetry breaking. One needs to establish if there is
another critical value $N_f^{c_2} < N_f^{c_1}$ which can
sufficiently dilute the gluon-gluon interactions to restore chiral
symmetry and deconfine color degrees of freedom. Such a phase
transition lies at the non perturbative boundary of the
interactions under scrutiny and hence we cannot expect to extract
sufficiently reliable information from  multiloop calculations of
the QCD $\beta$-function. Purely non perturbative techniques are
required to tackle the problem. Lattice studies in the infrared
indicate that just below $N_f^{c_1}$, chiral symmetry remains
unbroken and color degrees of freedom are
unconfined~\cite{Appelquist:2009}. Below this conformal window,
for an $8< N_f^{c_2} < 12$, the evolution of the beta function in
the infrared is such that QCD enters the phase of dynamical mass
generation as well as confinement.

Modern lattice analyses for this matter appear to strongly argue
in favour of a restoration for the chiral symmetric phase taking
place somewhere between $N_f \sim 8$ and $N_f \sim
10$~\cite{Iwasaki:2012,Aoki:2013}. In particular, the authors of
ref.~\cite{Aoki:2013}, with their study of the meson spectrum in
lattice QCD with eight light flavours using the Highly Improved
Staggered Quark action, gathered some striking evidences that
$N_f=8$ QCD still lies in the broken chiral symmetry phase but, at
the same time, suffering the effects from a remnant of the
infrared conformality (a large anomalous dimension for the quark
mass renormalization constant) indicating that the unbroken phase
is recovered near above $N_f \sim 8$. In the present work, we
intend to combine the Schwinger-Dyson machinery, well adjusted to
account for QCD phenomenology in the pion sector, with the latest
lattice data including twisted-mass dynamical light flavours in
order to provide a model for the chiral restoration mechanism, in
quantitative agreement with the above mentioned lattice studies.

\section{Chiral phase transition picture from Schwinger-Dyson
and Lattice gluon propagators}

In continuum, Schwinger-Dyson equations (SDEs) of QCD provide an
ideal framework to study its infrared properties,~\cite{SD:1949}.
These are the fundamental equations of any quantum field theory,
linking all its defining Green functions to each other through
intricately coupled nonlinear integral equations. As their formal
derivation through variational principle makes no appeal to the
weakness of the interaction strength, they naturally connect the
perturbative ultraviolet physics with its emerging non
perturbative properties in the infrared sector within the same
framework. The simplest two-point quark propagator is a basic
object to analyze dynamical chiral symmetry breaking and
confinement. Within the formalism of the SDEs, the inverse quark
propagator can be expressed as $
 S^{-1}(p) = {\cal Z}_2 (i \gamma \cdot p + m) +  \Sigma(p)
 $,
where $\Sigma(p)$ is the quark self energy:
 \bea
   \Sigma(p)= {\cal Z}_1 \int \frac{d^4q}{(2 \pi)^4} \, g^2 \Delta_{\mu
   \nu}(p-q) \frac{\lambda^a}{2} \gamma_{\mu} S(q)
   \Gamma_{\nu}^a(q,p) \;,
 \eea
where ${\cal Z}_1={\cal Z}_1(\mu^2,\Lambda^2)$ and ${\cal
Z}_2={\cal Z}_2(\mu^2,\Lambda^2)$ are the renormalization
constants associated respectively with the quark-gluon vertex and
the quark propagator. $\Lambda$ is the ultraviolet regulator and
$\mu$ is the renormalization point. The solution to this equation
is
\bea\label{eq:Sm1}
  S^{-1}(p) = \frac{i \gamma \cdot p + M(p^2)}{Z(p^2,\mu^2)} \ ,
\eeq
where $Z(p^2,\mu^2)$ is the quark wavefunction renormalization and
the quark mass function $M(p^2)$ is renormalization group
invariant. This equation involves the quark-gluon vertex
$\Gamma_{\nu}^a(q,p)$ and the gluon propagator $\Delta_{\mu
\nu}(p-q)$.

\subsection{Modelling the flavour behaviour for the gluon propagator}

\begin{figure}[h] 
{\centering
{\includegraphics[height=5cm,width=0.85\columnwidth]{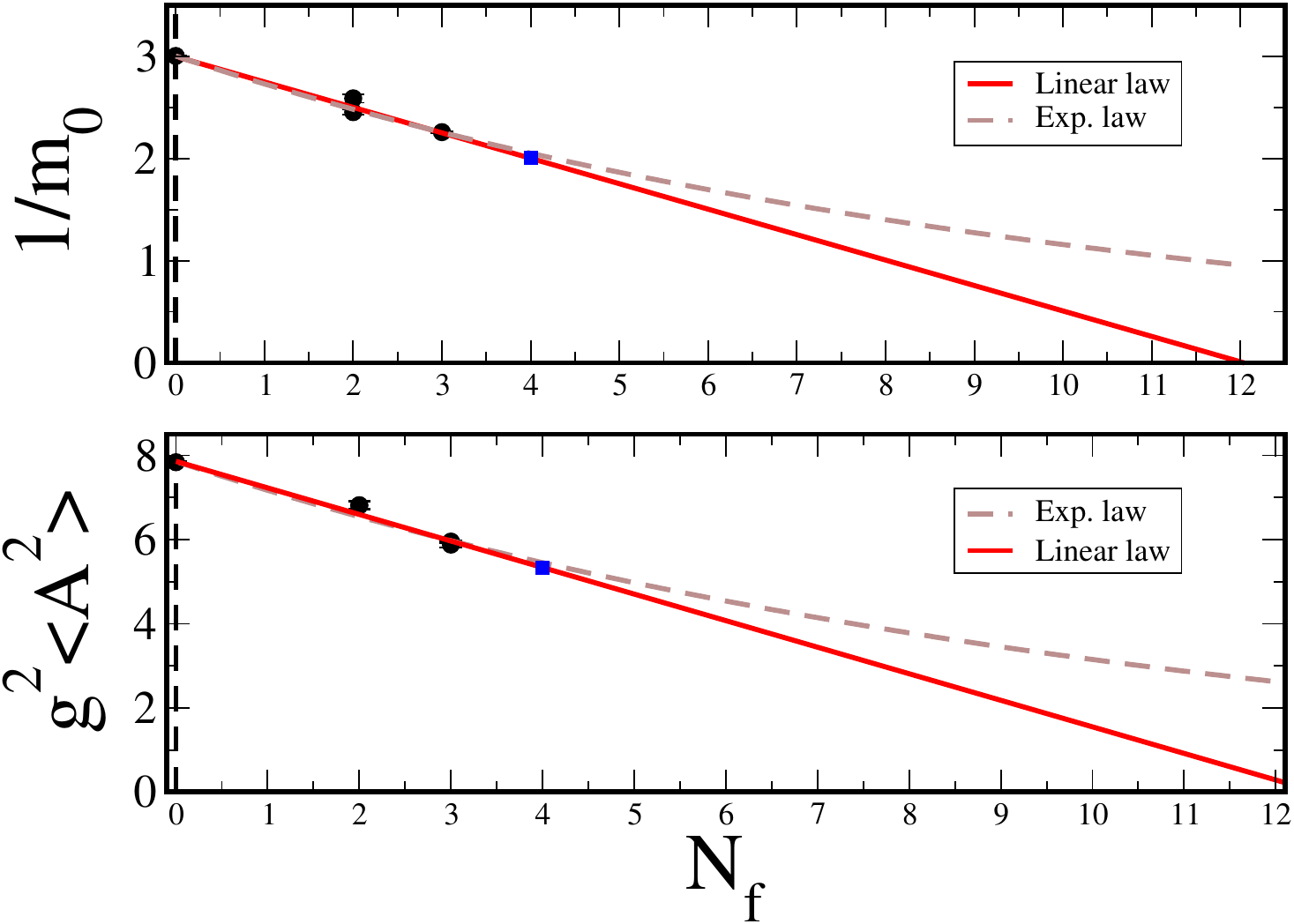}}
\par}
\caption{Parameters $g^2<A^2>$ and $1/m_0^2$ in terms of the
numbers of flavors and the fits with
Eqs.~(\ref{eq:masses},\ref{eq:massesexp}). The blue squares stand
for the extrapolated results at $N_f=$4 we used for
Fig.~\ref{fig:gluonprop2}.} \label{fig:parameters}
\end{figure}

\begin{figure}[h] 
{\centering
{\includegraphics[height=5cm,width=0.85\columnwidth]{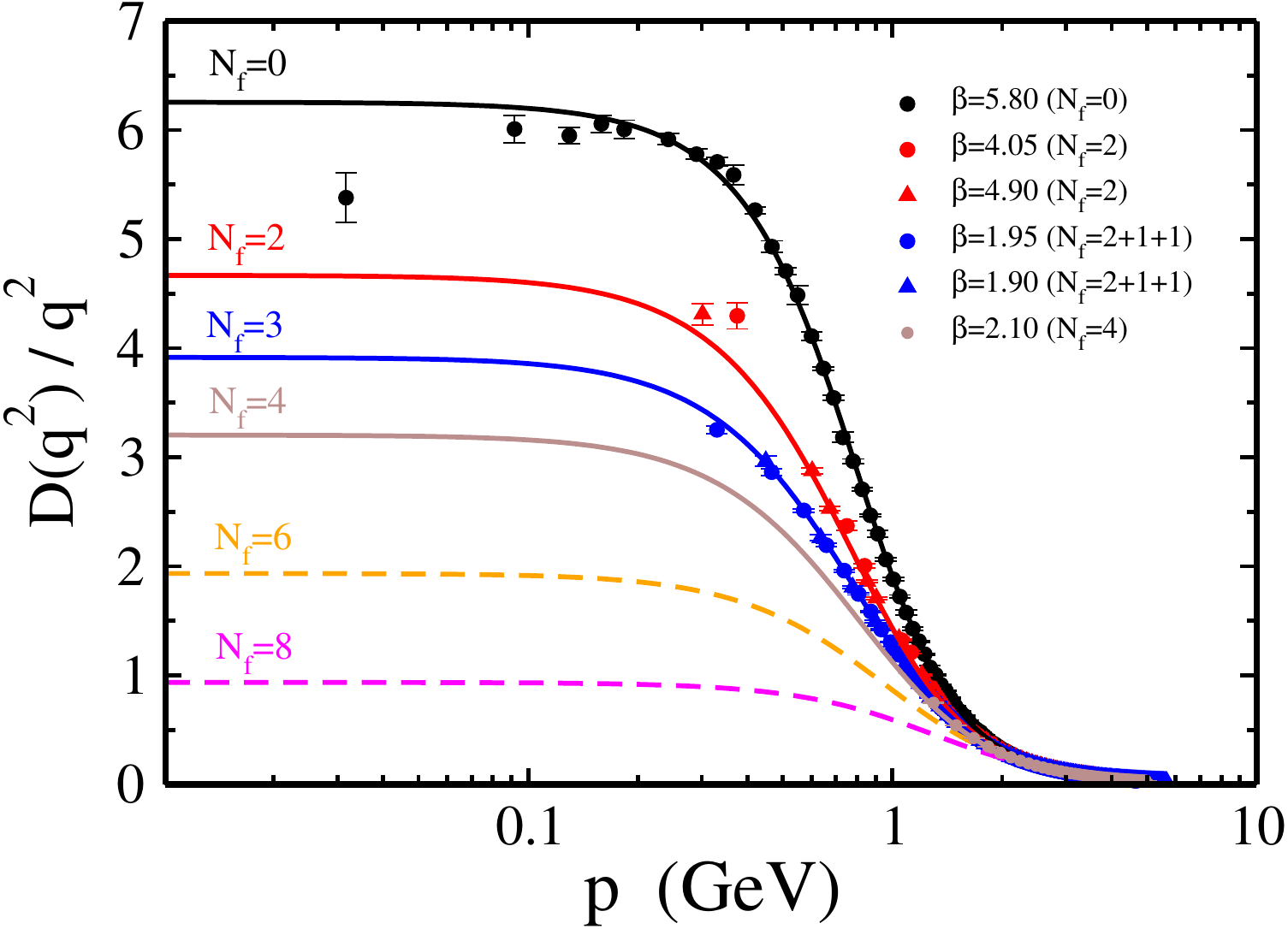}}
\par}
\caption{Lattice gluon propagator data in terms of momenta for
different number of fermion flavors and fits with \eq{eq:gluonR} and
the parameters of \eq{eq:masses}.} \label{fig:gluonprop}
\end{figure}

\begin{figure}[h] 
{\centering
{\includegraphics[height=5cm,width=0.85\columnwidth]{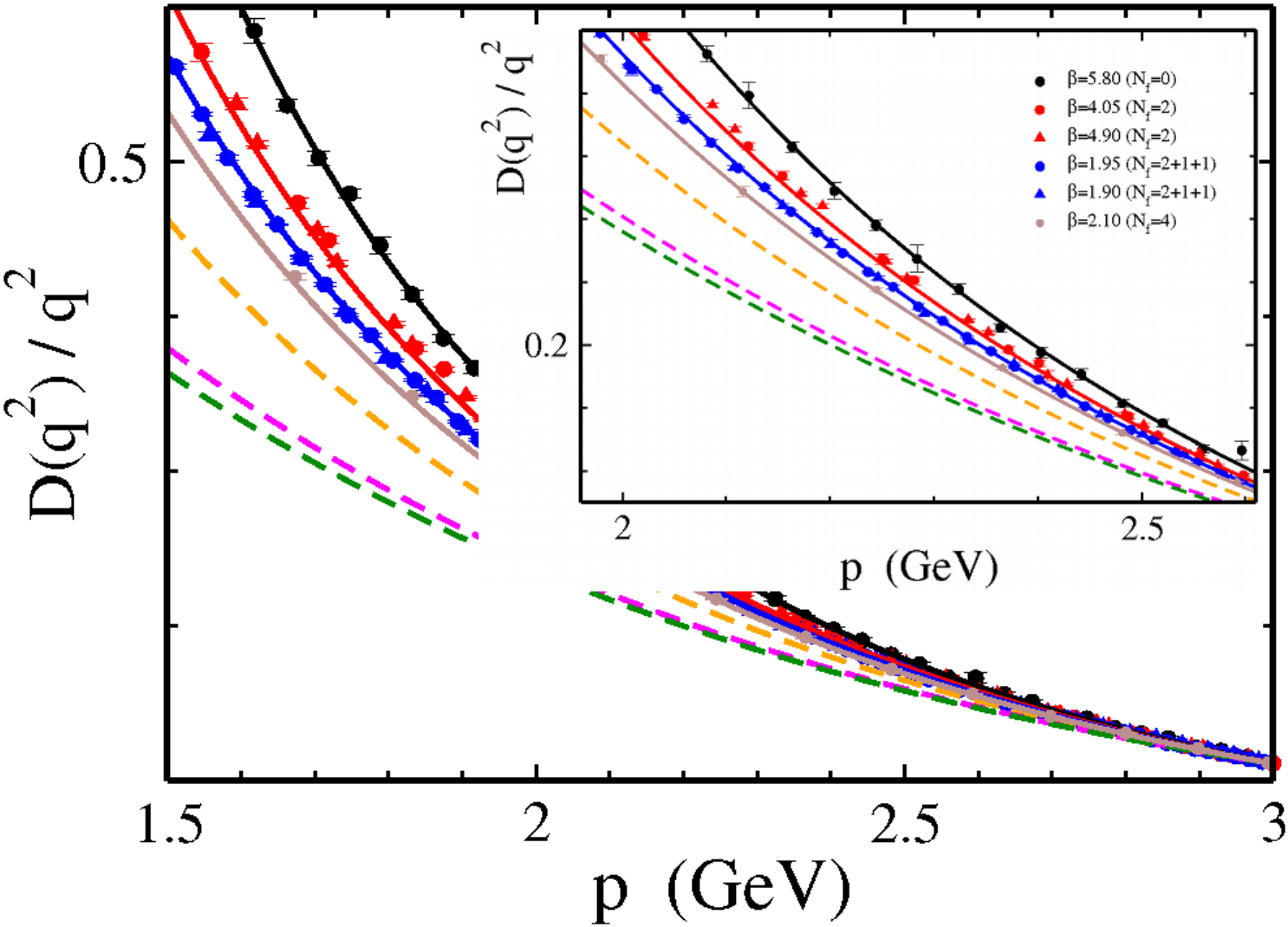}}
\par}
\caption{The same of Fig.~\ref{fig:gluonprop} but with the parameters of
set 2 and incorporating new small-volume lattice data for 4 degenerate fermion
flavors.}
\label{fig:gluonprop2}
\end{figure}

As a consequence of a patient effort, spanning several decades to
unravel gluon propagator $\Delta_{\mu \nu}$ in the infrared,
lattice as well as SDE studies have finally converged on its
massive or so called decoupling solution, see for
example~\cite{Gluon:2009}. After the gluon propagator solution in
the quenched approximation has been chiselled, we now have the
first quantitatively reliable glimpses of its quark flavor
dependence by incorporating $N_f=0,2$ light dynamical quark
flavors~\footnote{The dynamical flavors have been generated,
within the framework of ETM
collaboration~\cite{Baron:2010bv,Baron:2011sf,Blossier:2010ky,Blossier:2011tf},
with the mass-twisted lattice action~\cite{Frezzotti:2000nk},
while $N_f=0$ data have been borrowed
from~\cite{Bogolubsky:2007ud}.} and 2+1+1 (2 light degenerate
quarks, with masses ranging from 20 to 50 [MeV], and two non
degenerate flavors for the strange and the charm quarks, with
their respective masses set to 95 [MeV] and 1.51
[GeV])~\cite{Ayala:2012}. As we demonstrate shortly, in this last
2+1+1 case, the number of light quarks corresponds effectively to
3. This is exactly the result derived from the recently developed
{\it "partially unquenched"} approach to incorporate flavor
effects in the gluon SDE,~\cite{Aguilar:2012rz}. Their work is in
agreement with one of ~\cite{Ayala:2012} when the charm flavor is
assumed to decouple from gluons. In any case, this two-point
function serves as a crucial input to study the quark propagator.
The only other ingredient is the three-point quark-gluon vertex
$\Gamma_{\nu}^a(q,p)$. Significant advances have been made in
pinning it down through its key attributes in the ultraviolet and
infrared domains~\cite{Vertex:All}. More recently, the seeds of
the most general ansatz for the fermion-boson vertex appeared
in~\cite{Sanchez:2011} and its full blown extension was presented
in~\cite{Bermudez:2012}. Significantly, this ansatz contains
nontrivial factors associated with those tensors whose appearance
is expressly driven by dynamical chiral symmetry breaking in a
perturbatively massless theory. This novel feature enables a
direct and positive comparison with the best available symmetry
preserving solutions of the inhomogeneous Bethe-Salpeter equation
for the vector vertex. This encouraging outcome indicates that
this model is likely to provide a much-needed tool for use in
Poincar\'{e}-covariant symmetry-preserving studies of hadron
electromagnetic form factors. Furthermore, given the general
nature of constraints and the simplicity of the construction, a
straightforward extension of this approach is expected to yield an
ansatz adequate to the task of representing the
dressed-quark-gluon vertex. Before this is achieved, we restrict
ourselves to an efficacious approach. Following the lead of Maris
{\em et. al.}~\cite{Maris:1998}, we employ the following suitable
ansatz which has sufficient integrated strength in the infrared to
achieve dynamical mass generation:
  \bea
  {\cal Z}_1 g^2 \Delta_{\mu \nu}(p-q) \Gamma_{\nu}(p,q) \rightarrow g^2_{\rm eff}(q^2) \; \Delta^N_{\mu \nu}(p-q)
  \frac{\lambda^a}{2}
  \gamma_{\nu}\;,
 \eea
 where
 \bea
  \Delta^N_{\mu \nu}(q) &=& \frac{D(q^2)}{q^2} \; \left[ \delta_{\mu \nu} - \frac{q_{\mu} q_{\nu}}{q^2}
  \right] \; .
 \eea
The effective coupling $g_{\rm eff}$ is chosen to correctly
reproduce the static as well as dynamic properties of mesons below
1 GeV and reproduce perturbation theory in the ultraviolet, see
for example review~\cite{Review:2012} and references therein.
Moreover, our modern understanding of the flavor dependence of the
gluon propagator provides us with the solid basis to use the
following non perturbative model~\cite{Dudal:2012zx}:
 \beq\label{eq:gluonR} D(q^2) \ = \ \frac{z(\mu^2) \ q^2
 (q^2+M^2)}{\displaystyle q^4 + q^2 \left(M^2-{13} g^2 \VA /24
 \right) + M^2 m_0^2}
 \eeq
to describe the gluon dressing renormalized in MOM scheme at
$q^2=\mu^2$. This model is based on the tree-level gluon
propagator obtained with the Refined Gribov-Zwanziger (RGZ)
action~\cite{Dudal:2008sp} which has been shown to describe
properly the lattice data in the infrared sector (see
refs.~\cite{Dudal:2010tf,Dudal:2012zx}). The overall factor
$z(\mu^2)$ is introduced to guarantee the multiplicative MOM
renormalization prescription, namely, $D(\mu^2) = 1$, and implies
no physical consequence as the effective coupling, $g_{\rm
eff}(q^2)$, is further adjusted to reproduce properly the meson
phenomenology. We obtain the mass parameters of \eq{eq:gluonR} by
fitting it to the gluon propagator lattice data analyzed in
Ref.~\cite{Ayala:2012}. $M^2$ is related to the condensate of
auxiliary fields, emerging merely to preserve locality for the RGZ
action. A free fit of the lattice data suggests that it does not
depend on the number of fermion flavors (we find $M^2=4.85$
[GeV$^2$]). Dimension two gluon condensate
$\VA$,~\cite{Boucaud:2000nd}, and $m_0^2 \ = z(\mu^2) \lim_{q^2\to
0} {q^2}/{D(q^2)}$ are flavor dependent and we look for their best
fits. In order to cover a wide range of possibilities within
reason, we assume their evolution with the flavor number to be
driven either by a simple linear scaling law
\beq\label{eq:masses}
 m_0^{-1}(N_f) &=& {m_0^{-1}(0)} \ (1 - A N_f)  \nonumber    \\
g^2 \VA(N_f) &=& g^2\VA(0) \ (1 - B N_f)  \ ,
\eeq
as data appear to suggest, or by an exponential law
\beq\label{eq:massesexp}
 {m_0^{-1}(N_f)} &=& {m_0^{-1}(0)} \ e^{-A N_f}  \nonumber \\
g^2 \VA(N_f) &=& g^2\VA(0)  \ e^{-B N_f} \ ,
\eeq
which allows for the possibility that the gluon propagator becomes
infinitely massive only when the number of light quark flavors
tends to infinity. The best-fit of the $m_0$ and $g^2\VA$ from
lattice data will require $m_0(0)=0.333$ GeV and
$g^2\VA(0)=7.856$, in both cases, $A=0.083$ and $B=0.080$, for the
linear case, and $A=0.095$ and $B=0.091$, for the exponential one.
Eq.~(\ref{eq:gluonR}) now provides a prediction for the gluon
propagator for arbitrarily large $N_f$, see
Fig.~\ref{fig:parameters}, while Fig.~\ref{fig:gluonprop} shows
the corresponding gluon propagator along with the lattice data
superimposed~\cite{Ayala:2012}. We also include some very recent
gluon propagator data obtained from lattice simulations with four
degenerate light twisted-mass flavors~\footnote{The gluon
propagator lattice data for 4 light flavors have been taken from
ETMC~\cite{ETMCNf4}. Simulated at small volumes, they are only
available for momenta above 1.25 GeV and hardly allow for a fit
with \eq{eq:gluonR}. Nevertheless, they can be used to check our
modelling of the flavor evolution.}. These new data are rather
well described by \eq{eq:gluonR} evaluated for the mass parameters
extrapolated to $N_f$=4 with \eq{eq:masses} (see the zoomed plot
in Fig.~\ref{fig:gluonprop2}). This observation strongly supports
that $N_f=$2+1+1 gluon data indeed correspond to three light
flavors.


Thus, we can efficaciously model the dilution of the gluon-gluon
interactions with increasing flavor number in order to study the
chiral restoration mechanism. We can now employ the gap equation
to provide quantitative details of chiral symmetry breaking in
terms of the quark mass function for an increasing number of light
quarks.

\subsection{Results}

In the following, we mostly discuss the results obtained by
employing the linear law and state the effect of exponential
extrapolation afterwards. We take the effective coupling $g_{\rm
eff}(q^2)$ to be independent of $N_f$ which is justified by the
results of~\cite{Ayala:2012} (see Eq.(5.2)) which suggest that an
effective coupling can be constructed such that there is an
absence of any flavor dependence in the infrared region, more
precisely starting from $q^2 \lesssim 1$ GeV$^2$.
\begin{figure}[h] 
{\centering {\includegraphics[width=0.9\columnwidth]{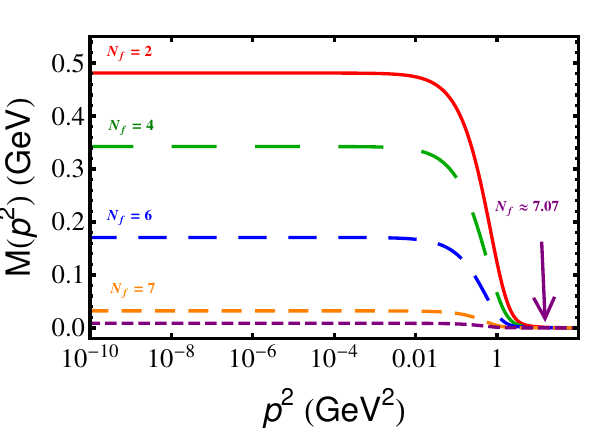}}
\par}
\caption{The quark mass function diminishes in height for
increasing light quark flavors (here with \eq{eq:masses}). Above
$N_f \approx 7.07$, only the chirally
symmetric 
solution exists.}\label{fig:quarkmass}
\end{figure}
Note that we have not considered the flavor dependence which would
arise from the quark-gluon vertex. No explicit handle on this
dependence is available at the moment. Within the Abelian theory
of QED, restrictions imposed by the all order multiplicative
renormalizability of the photon propagator may provide a handle on
the transverse part of the electron-photon vertex, see the last
reference in~\cite{Vertex:All}. A consequent non perturbative
construction of such a vertex with imprints of the massless
charged fermion flavors and its subsequent extension to QCD is
still not available.
\begin{figure}[h] 
{\centering {\includegraphics[width=0.9\columnwidth]{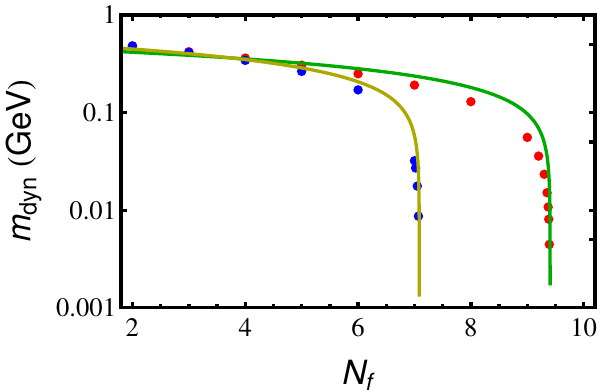}}
\par}
\caption{Quark pole mass in the Euclidean space clearly
demonstrates that chiral symmetry is restored above a critical
number of quark flavors. Blue (red) points correspond to linear (exponential) case.
The solid line is the mean-field scaling, Eq.~(\ref{scaling}).} \label{fig:polemass}
\end{figure}
Once the quark mass function is available for varying light quark
flavors (see Fig.~\ref{fig:quarkmass} for the linear case), one can investigate any of the interrelated order
parameters, namely, the Euclidean pole mass defined as $m_{\rm
dyn}^2 + M^2(p^2= m_{\rm dyn}^2)= 0$, the quark-antiquark
condensate which is obtained from the trace of the quark
propagator or the pion leptonic decay constant $f_{\pi}$ defined
through the Pagel-Stokar equation~\cite{Pagel:1979}, or through
considering the residue at the pion pole of the meson propagator.
Each of these quantities involves the quark wave-function
renormalization, the mass function and/or its derivatives and is
hence calculable from the solution for the full quark propagator.
Moreover, these order parameters can help locate the critical
number of flavors above which chiral symmetry is restored.

We investigate these three order parameters and choose to present
here the Euclidean pole mass of the quark in
Fig.~\ref{fig:polemass} for the linear (exponential) case and show
that, at a critical value of about $N_f^c \approx 7.1$ ($N_f^c
\approx 9.4$), chiral symmetry appears restored. The phase
transition appears second order, described by the following mean
field behavior (solid lines in Fig.~\ref{fig:polemass})~:
 \bea
 m_{\rm dyn} \thicksim \sqrt{N_f^{c_2} - N_f} \;. \label{scaling}
 \eea
This behavior of QCD resembles that of the toy version of QED with
large electromagnetic coupling with or without the inclusion of
4-fermion operators to render the theory closed~\cite{QEDNf}.

\begin{figure}[h] 
{\centering {\includegraphics[width=0.9\columnwidth]{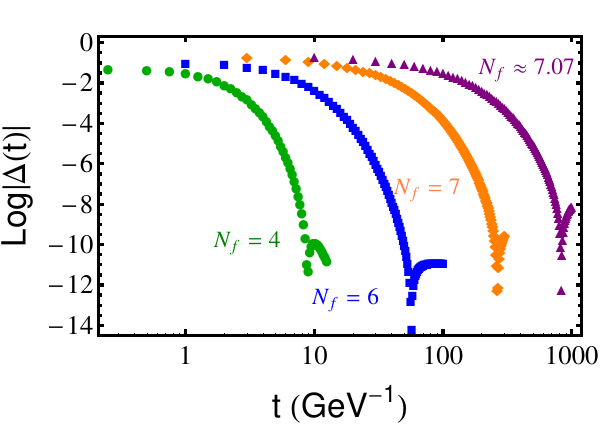}}
\par}
\caption{Spatially averaged Euclidean space 2-point Schwinger
function $\Delta(t)$ develops oscillations for large times which
corresponds to the non-existence of asymptotically stable free
quark states. For sufficiently large values of $N_f$, the first
minimum of these oscillations is pushed all the way to infinity,
thus ensuring the existence of a pole on the time-like axis, a
property of free particle propagators.} \label{oscillations}
\end{figure}
It has been established that confinement is related to the
analytic properties of QCD Schwinger functions which are the
Euclidean space Green functions, namely, propagators and vertices.
One deduces from the reconstruction
theorem~\cite{Reconstruction:1980s} that the only Schwinger
functions which can be associated with expectation values in the
Hilbert space of observables; namely, the set of measurable
expectation values, are those that satisfy the axiom of reflection
positivity. When that happens, the real-axis mass-pole splits,
moving into pairs of complex conjugate singularities. No
mass-shell can be associated with a particle whose propagator
exhibits such singularity structure. We  define the following
Schwinger function:
\begin{figure}[h] 
{\centering {\includegraphics[width=0.9\columnwidth]{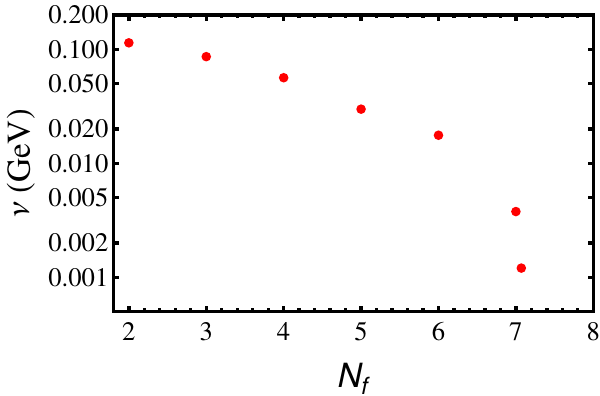}}
\par}
\caption{The order parameter for confinement $\nu (N_f) = 1 /
\tau_1(N_f)$, where $\tau_1(N_f)$ is the location of the first
zero of~Eq.~(\ref{Deltat}). Comparison with
Fig.~\ref{fig:polemass} suggests that quarks get deconfined when
chiral symmetry is restored.} \label{critconf}
\end{figure}
 \bea
 \Delta(t) &=& \int d^3x \int \frac{d^4p}{(2 \pi)^4} \;
 {\rm e}^{i(p_4t+ {\bf{p}} \cdot {\bf x} ) } \sigma_s(p^2) \ ,
 \label{Deltat}
 \eea
 to study the analytic properties of the quark propagator; where
 $\sigma_s(p^2)$ is the scalar term for the quark propagator in
 Eq.~(\ref{eq:Sm1}), that can be written in terms of the quark wavefunction renormalization and mass function as $Z(p^2,\mu^2)
 M(p^2)/(p^2+M(p^2))$. One can show that if there is a stable asymptotic state
 associated with this propagator, with a mass $m$, then
 $\Delta(t) \thicksim {\rm e}^{-mt}$, whereas two complex
 conjugate mass-like singularities, with complex masses
 $\mu = a \pm i b$ lead to an oscillating behavior of the sort
 $\Delta(t) \thicksim {\rm e}^{-at} {\rm cos}(b t + \delta)$ for large $t$,~\cite{Maris:1995}.
 Fig.~\ref{oscillations} analyzes this function for varying
 $N_f$, in the linear extrapolation case. The existence of oscillations clearly demonstrates that
 the quarks correspond to a confined excitation for small $N_f$. With increasing
 $N_f$, the onslaught of oscillations moves towards higher values
 of $t$ and eventually never takes place above a critical $N_f$
 when quarks deconfine and correspond to a stable asymptotic
 state.
As an order parameter of confinement, we therefore employ $\nu
(N_f) = 1 / \tau_1(N_f)$, where $\tau_1(N_f)$ is the location of
the first singularity,~\cite{Sanchez:2009}. The first oscillation is pushed to
infinity when confinement is lost. It is notable that when the dynamically
generated mass approaches zero, $\nu (N_f)$ diminishes rapidly
(see~Fig.~\ref{critconf}). This highlights the intimate
connection between chiral symmetry restoration and deconfinement.
In fact, within our numerical accuracy, $N_f^c$ is found
to be the same for both the transitions.

The results with the exponential and linear flavor extrapolations are
qualitatively the same, leading to identical conclusions. They only
quantitatively differ by the critical flavor numbers, although
both are pretty much in the same ballpark: $N_f^c\simeq 7.1$ and
$N_f^c\simeq 9.4$. Note that both the parameterizations, so far apart as to
have an infinitely massive gluon at $N_f \approx 12$ or $N_f
\Rightarrow \infty$, restore chiral symmetry and trigger
deconfinement at so similar value of light quark flavors.

\section{Conclusions}

We have performed a Poincare-covariant SDE analysis of the latest
lattice results for the quark flavor dependence of the gluon
propagator in the infrared, provided then with a model for the
dilution of the gluon-gluon interaction with increasing number of
light quarks and finally with a picture for the chiral restoration
mechanism. The quantitative analysis, following this approach,
hints at the restoration of chiral symmetry and deconfinement in
QCD when the number of light quark flavors exceeds a critical
value of $N_f^{c_2} \thickapprox 8.2 \pm 1.2$. This is in perfect
agreement with the state-of-the-art lattice investigations of
chiral symmetry restoration in QCD~\cite{Iwasaki:2012,Aoki:2013}
and supports that the model presented here for the chiral
restoration mechanism is properly capturing the relevant physics
for the problem. Having said that, it will surely be illuminating
to incorporate and study the effect of the flavor dependent
quark-gluon vertex and, moreover, solve the coupled system of the
Green functions involved simultaneously. All this is for future.

\bigskip

\noindent {\bf Acknowledgments} We acknowledge D. Schaich for a fruitful communication.
This work was supported  by the grants: CIC, UMICH, Mexico, 4.10 and 4.22, CONACyT (Mexico)
82230 and 128534, and MINECO (Spain) research project
FPA2011-23781.

\end{document}